%% file: Main_EuMW_Paper.tex
\documentclass[conference]{IEEEtran}
\include{macros}

\usepackage{amsmath}% for double integral symbol in this template
\usepackage{times}% use times font for the paper instead of default Computer Modern fonts
\usepackage{graphicx}% for figures
\usepackage{multirow}% to allow multiple-row elements in tabular environment
\usepackage[none]{hyphenat}% turn off hyphenation to make text extraction and indexing easier
\usepackage{float}% better control of floating figures and tables
\usepackage[nolist]{acronym}
\usepackage{amssymb}
\usepackage{amscd}
\usepackage{amsmath}
\usepackage{graphicx}
\usepackage{epsfig}
\usepackage{amsthm}
\usepackage{graphics}
\usepackage{psfrag}
\usepackage{rotating}
\usepackage{amsmath} 
\usepackage{amsfonts}
\usepackage{url}
\usepackage{color}
\usepackage{epstopdf}
\usepackage{amsthm}
\usepackage{tikz}
\usepackage{mathtools}
\usepackage{multirow}
\usepackage[final]{pdfpages}
\usepackage{enumitem}
\usepackage{multicol}
\usepackage[nolist]{acronym}
\usepackage{siunitx}
\usepackage[font= small]{caption}
\usepackage{subcaption}
\usepackage[ruled]{algorithm2e}
\usepackage{ragged2e}
\usepackage{placeins}
\usepackage{array}
\usepackage{multirow}
\usepackage{pgfplots}
\usepackage{pgfplotstable}
\usepackage{calrsfs}
\pgfplotsset{compat=1.12}
\usepgfplotslibrary{polar}

\usepackage{hyperref}
\hypersetup{
    colorlinks=true,
    linkcolor=blue,
    filecolor=magenta,      
    urlcolor=blue,
    }
\begin{document}
%%%%%%%%%%%%%%%%%%%%%%%%%%%%%%%%%%%%%%%%%%%%%%%%%%%%%%%%%%%%%%%%%%%%%%%%%%%%%
% We use \raggedbottom to avoid latex adding vertical space around headings.
% This gives a better idea to the author about how much white space remains
% as the page limit is approached.
\raggedbottom
%
%%%%%%%%%%%%%%%%%%%%%%%%%%%%%%%%%%%%%%%%%%%%%%%%%%%%%%%%%%%%%%%%%%%%%%%%%%%%%
% PAPER TITLE AND AUTHOR BLOCK
%
% The paper title can use linebreaks \\ within to get better formatting if desired.
%
\title{Extended Target Parameter Estimation and Tracking with an HDA Setup for ISAC Applications}
%
% Next we define the author names and affiliations.
% Author names are listed using \IEEEauthorblockN{} with comma separators between names.
% Affiliations are listed using \IEEEauthorblock{} with \\ separators between affiliations.
% Symbols marking author-affiliation relations are output using \EuMWauthorrefmark{}.
% At the end of the affiliation list is the list of author emails.
% See below for examples of each of these.
%
\author{%
\IEEEauthorblockN{%
% Fernando Pedraza\EuMWauthorrefmark{\#1},
% Saeid K. Dehkordi\EuMWauthorrefmark{\#2},  
% Jan Hauffen\EuMWauthorrefmark{\#},
% Peter Jung\EuMWauthorrefmark{\#},
% Giuseppe Caire\EuMWauthorrefmark{\#}
Fernando Pedraza,
Saeid K. Dehkordi, 
Jan C. Hauffen,
Shuangyang Li, 
Peter Jung, and
Giuseppe Caire
}% \IEEEauthorblockN Names
\IEEEauthorblockA{%
Technical University of Berlin, Germany\\
\{f.pedrazanieto, s.khalilidehkordi, j.hauffen, shuangyang.li, peter.jung, caire\}@tu-berlin.de\\
}% \IEEEauthorblockA Affils
}% \author
%
% Next we make the title/author block using the information defined above.
\maketitle

\begin{acronym}
	\acro{AWGN}{additive white Gaussian noise}
	\acro{MIMO}{multiple-input multiple-output}
	\acro{OTFS}{orthogonal time frequency space}
	\acro{ISAC}{integrated sensing and communication}
    \acro{JSAC}{joint sensing and communication}
	\acro{sTHZ}{sub-THz}
	\acro{SNR}{signal-to-noise ratio}
	\acro{mmWave}{millimeter wave}
	\acro{ML}{maximum likelihood}
	\acro{V2X}{vehicle-to-everything}
	\acro{OFDM}{orthogonal frequency division multiplexing}
	\acro{FMCW}{frequency modulated continuous wave}
	\acro{LoS}{line-of-sight}
	\acro{ISFFT}{inverse symplectic finite Fourier transform}
	\acro{SFFT}{symplectic finite Fourier transform}
	\acro{HPBW}{half-power beamwidth}
	\acro{ULA}{uniform linear array}
	\acro{CRLB}{Cram\'er-Rao lower bound}
	\acro{RF}{radio frequency}
	\acro{BF}{beamforming}
	\acro{RMSE}{root MSE}
	\acro{AoA}{angle of arrival}
	\acro{ISI}{inter-symbol interference}
	\acro{SI}{self-interference}
	\acro{TDD}{time division duplex}
	\acro{Tx}{transmitter}
	\acro{Rx}{receiver}
	\acro{SIC}{successive interference cancellation}
	\acro{PD}{probability of detection}
	\acro{HDA}{hybrid digital-analog}
	\acro{PSD}{power spectral density}
	\acro{FWHM}{full width at half maximum}
	\acro{SLL}{side lobe level}
	\acro{BS}{base station}
    \acro{FoV}{field of view}
    \acro{CFAR}{constant false alarm rate}
    \acro{OS-CFAR}{ordered statistic constant false alarm rate}
    \acro{PSLR}{peak-to-sidelobe ratio}
    \acro{DFT}{discrete Fourier transform}
    \acro{MSP}{multi-scatter-point}
    \acro{UE}{user equipment}
    \acro{ISL}{integrated side-lobe Level}
    \acro{PSL}{peak side-lobe level}
    \acro{CAV}{connected automated vehicle}
    \acro{DSRC}{dedicated short-range communication}
    \acro{V2X}{vehicle-to-everything}
    \acro{V2V}{vehicle-to-vehicle}
    \acro{LFM}{linear frequency modulation}
    \acro{SC} {single carrier}
    \acro{STF}{short training field}
    \acro{CEF}{channel estimation field}
    \acro{HDA-AF}{hybrid digital-analog ambiguity function}
    \acro{PA}{phased array}
    \acro{CPI}{coherent processing interval}
    \acro{CSI}{channel state information}
    \acro{BT}{beam tracking}
    \acro{Tx}{Transmitter}
    \acro{Rx}{Receiver}
    \acro{CDF}{cumulative distribution function}
\end{acronym}

%%%%%%%%%%%%%%%%%%%%%%%%%%%%%%%%%%%%%%%%%%%%%%%%%%%%%%%%%%%%%%%%%%%%%%%%%%%%%
% ABSTRACT paragraph.
% in the abstract paragraph.
%
\begin{abstract}
We investigate radar parameter estimation and beam tracking with a \ac{HDA} architecture in a multi-block measurement framework using an extended target model. In the considered setup, the backscattered data signal is utilized to predict the user position in the next time slots. Specifically, a simplified maximum likelihood framework is adopted for parameter estimation, based on which a simple tracking scheme is also developed. 
Furthermore, the proposed framework supports adaptive transmitter beamwidth selection, whose effects on the communication performance are also studied.
Finally, we verify the effectiveness of the proposed framework via numerical simulations over complex motion patterns that emulate a realistic \ac{ISAC} scenario.   
\end{abstract}
\begin{IEEEkeywords}
ISAC, beam tracking, extended target tracking, hybrid digital-analog.
\end{IEEEkeywords}
%
%%%%%%%%%%%%%%%%%%%%%%%%%%%%%%%%%%%%%%%%%%%%%%%%%%%%%%%%%%%%%%%%%%%%%%%%%%%%%
% THE REST OF THE PAPER follows.
%

\section{Introduction}
To accommodate the needs of many emerging applications, such as autonomous vehicles and extended reality, 6G wireless networks are envisioned to provide not only high-quality communications but also highly accurate and robust sensing services~\cite{SP-ISAC-Heath-Liu}. Therefore, \ac{ISAC} has been widely acknowledged as one of the key components in 6G wireless networks~\cite{Masouros_Survey}. 

In this work, we specifically focus on automotive applications of \ac{ISAC}, where a multi-antenna \ac{Tx} unit, e.g. a \ac{BS} as a roadside infrastructure, communicates with vehicles in the traffic scenario over a doubly-dispersive channel. 
Given that \ac{ISAC} applications are expected to operate in the \ac{mmWave}  regime~\cite{SP-ISAC-Heath-Liu}, known to be sparse in the beam-space domain \cite{mmWaveChannel-Hanzo}, accurate \ac{CSI} is required for efficient communications. Acquiring this information leads to overhead, which can become prohibitively large when considering high mobility scenarios. Given the temporally correlated structure of the \ac{CSI}, past observations can be used to simplify the estimation process, a technique we refer to as \ac{BT}.

%A major bottleneck in the practical design of \ac{mmWave} \ac{ISAC} systems is the excessive energy required to process, transmit and sample large dimensional signals. As a consequence, we focus on \ac{HDA} antenna architectures, such that the spatial dimension is reduced when transforming from the analog into the digital domain \cite{HDA_Sohrabi}. 
% therefore dynamically inferring the channel state given past observations (commonly referred \textit{beam tracking}) can provide a viable solution to the problem at hand. 

% Additionally, the large bandwidths of \ac{mmWave} systems, and typically large antenna arrays to provide high beamforming gain, motivate the use of energy efficient and low hardware complexity \ac{HDA} architectures \cite{HDA_Sohrabi}. 

\ac{ISAC} for automotive applications considering the above is not a new topic (more details can be found in~\cite{Cheng2022ISAC_VCN,DuZhen2023} and the references therein). However, most of the existing works have focused on the point target model, which greatly simplifies the problem of parameter estimation and BT. The point target model may not accurately reflect the considered problem in practice, due to the fact that users, such as motorcycles, bicycles, and vehicles, are typically observed under the \textit{extended target} model by radar sensors \cite{Thoma}. Another major bottleneck in the practical design of \ac{mmWave} \ac{ISAC} systems is the excessive energy required to process, transmit and sample large dimensional signals, a consequence of the wide bandwidths required to achieve high data rates and enhance range resolution, as well as the large number of spatial degrees of freedom required to perform angular estimation and apply spatial multiplexing. As a consequence, we focus on \ac{HDA} antenna architectures, such that the spatial dimension is reduced when transforming from the analog into the digital domain \cite{HDA_Sohrabi}. 
% . Additionally, the large bandwidths of \ac{mmWave} systems, and typically large antenna arrays to provide high beamforming gain, motivate the use of energy efficient and low hardware complexity \ac{HDA} architectures \cite{HDA_Sohrabi}. 

In this work, we model the user as an extended target for the radar. The previous work in \cite{Liu_Caire} has considered target tracking for an extended target model in the \ac{ISAC} framework, however, the work has assumed that the scatter points can be recovered via some matched filtering techniques in a system with very high range and Doppler resolutions and considered a fully digital array. Furthermore, the mentioned work does not consider any concrete modulation scheme. 
Against the existing literature, we propose a novel framework of ISAC for automotive applications in this paper, which adopts the extended target model with more practical settings and provides an energy efficient solution by focusing on \ac{HDA} antenna architectures.
% , such that the spatial dimension is reduced when transforming from the analog into the digital domain \cite{HDA_Sohrabi}. 
Specifically, in the considered setup, following initial target (user) detection (see \cite{OTFS_MIMO_AV_22} for details), an \ac{OFDM} modulated communication signal stream is transmitted to each acquired user via a dedicated radio frequency (RF) chain, where the \ac{Tx}  beamwidths can be selected adaptively according to the predicted user locations. The radar receiver then takes advantage of the backscattered communication signal to estimate the user parameters (angle of arrival, range, and Doppler velocity). The major contributions of the paper are summarized below:

\begin{itemize}
\item With the proposed framework, we demonstrate that, by using hardware and energy-efficient array architectures it is possible to reliably establish a communication link with a user traversing complex trajectories. Particularly, the proposed framework does not rely on models describing the mobility patterns of users, making our approach valid in general scenarios. 
\item By concentrating on extended targets, our framework is more realistic than the commonly used point target assumption, especially in vehicular \ac{ISAC} scenarios with large bandwidths and antenna arrays.
\item We propose a simple scheme to adapt the beamwidth at the \ac{Tx} based on the estimated position of the target. We demonstrate the advantages of our adaptive beamwidth method, including a more uniform coverage of the area served by the \ac{BS}.
% {\color{red}We shall add some based on the adaptive beamwidth.}
\end{itemize}

% \par The main contribution of this work is to demonstrate that, by using hardware and energy-efficient array architectures it is possible to reliably establish a communication link with a user traversing complex trajectories. Differently from most current works, the \ac{BS} does not rely on models describing the mobility patterns of users, making our approach valid in general scenarios. Furthermore, by concentrating on extended targets, our modeling is more realistic than the commonly used point target assumption, especially in vehicular \ac{ISAC} scenarios with large bandwidths and antenna arrays.

%%%%%%%%%%%%%%%%%%%%%%%%%%%%%%%%%%%%%%%%%%%%%%%%%%%%%%%%%%%%%%%%%%%%%%%%%%%%%

\section{System and Signal Model}\label{sec:sys_model}
The considered \ac{ISAC} system operates in a \ac{mmWave} channel with carrier frequency $f_c$ and bandwidth $W$ sufficiently smaller than $f_c$, such that the narrow-band array response assumptions hold. The \ac{BS} transmitter and the radar receiver are co-located. For simplicity, we assume that the Tx array and the Rx radar array coincide and that the Tx and Rx signals are separated via full-duplex processing \cite{duarte2010fullduplexWireless}. Note that the co-located setup implies that the angles of departure and arrival coincide. Aiming at hardware cost and energy efficiency, we consider a \textit{Fully-Connected} \ac{HDA} array architecture (see e.g.\cite{HDA_Sohrabi}) where the \ac{BS} is equipped with $\Nrf$ Tx RF chains connected to an antenna array with $\Na$ elements. For communication, the \ac{BS} tranmits $\Ns \geq 1$ data streams through a beamforming matrix $\Fm = [\fv_1, \dots, \fv_{\Ns}]$ where $\fv_q~,~q \in [1:\Ns]$ denotes the $q$-th column of $\Fm$ associated to the $q$-th data stream. We have designed the \ac{Tx} beamformers $\fv$ such that each covers a relatively narrow section of the beam-space with a constant gain, and very low gain elsewhere (see \cite{OTFS_MIMO_AV_22} for details), such that $\fv^\herm_{q}\fv_{q'} \approx 0~,~\forall q'\neq q$. The backscattered signal from the user is then used for radar processing. Note that, by  estimating the user parameters we aim to eliminate the need for active feedback in uplink. In the following, a single data stream (i.e. $\Fm=\fv$) is pointed toward the user and radar parameter estimation in a \textit{multi-block} framework is carried out. This is necessary since the \ac{HDA} architecture does not allow conventional MIMO radar processing. To this end, the Rx \ac{BF} matrices vary from block to block. Define a codebook given by a set of $D>\Nrf$, \ac{DFT} orthogonal beams as $\mathcal{U}_{\text{DFT}} \coloneqq \{\uv_{1},...,\uv_{D}\}$ selected from the Fourier basis of dimension $\Na$, such that they cover a desired region of interest (i.e. covering the illuminated region by the Tx) in the beam space (see Fig.~\ref{fig:system_scheme}. At each block $b\in \{0, \dots, B-1\}$, $\Nrf$ beams are selected from $\mathcal{U}_{\rm DFT}$ at random. Such a scheme can easily be justified as a result of insignificant target movement within the time interval of $B$ \ac{OFDM} blocks to acquire the signal, where B is typically small (refer to Section~\ref{sec:numeric_res}). 
%%%%%%%%%%%%%%%%%%%%%%%%%%%%%%%%%%%%%%%%%%%%%%%%%%%%%%%%%%%%%%%%%%%%%%%%%%%%%
\subsection{Channel model}
\begin{figure}
	\centering
	%\hspace{3cm}
    \hspace*{-1em} 
	\includegraphics[width=0.99\linewidth]{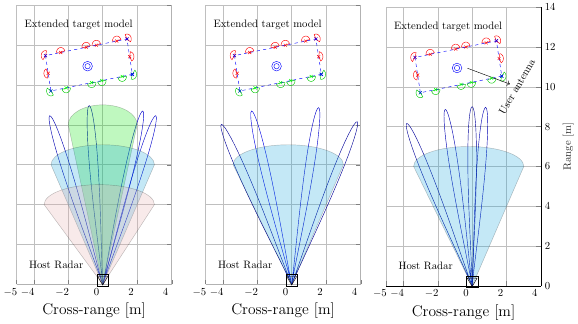}
	\caption{Extended target model and multi-block reception. The transparent beam indicates the illuminated region of the Tx beam, while the Rx beams vary over the 3 blocks. The different colors of the scatter points on the target indicate illumination with respect to the radar and change based on positioning. The left-most figure depicts the variable-width \ac{Tx} beamformers discussed in Section \ref{BF_arch}.}
	\label{fig:system_scheme}
\end{figure}

    We consider an extended target model, which represents a physically expansive \textit{user} equipment, such as an automobile (see Fig.~\ref{fig:system_scheme}). The applied model takes into account the viewing angle of the vehicle as observed by the radar, such that the number of scattering points from the vehicle dynamically changes along the trajectory.  Assuming the target to be composed of $P$ point scatterers, we adopt the widely used mmWave radar channel model (see e.g.  \cite{SP-ISAC-Heath-Liu}) where the received echo is the superposition of  all scatterers. The impulse response of this channel is given by:
    \begin{align}\label{eq:channel}
        \Hm (t, \tau) =  \sum_{p=0}^{P-1}h_{p} \av(\phi_{p}) \av^H(\phi_{p})\delta(\tau-\tau_{p})  e^{j2\pi \nu_{p} t}\,,
    \end{align}
    where $h_{p}$, $\tau_{p}$, $\nu_{p}$ and $\phi_{p}$ are respectively the complex channel coefficient, delay, Doppler and angle of arrival (AoA) of the $p$-th scatterer.  For simplicity, we focus on uniform linear arrays (ULA), such that their array response vectors have elements given by $[\av(\phi_{p})]_{i} = e^{j\pi(i-1)\sin(\phi_{p})},\quad i\in0,\dots,\Na-1$.
    The channel coefficient $h_{p}$, is given by the radar equation and satisfies $|{h_{p}}|^{2} = \frac{\lambda^{2}\sigma_{{\rm rcs}, p}}{(4\pi)^{3}d^{4}_{p}}$ where $\lambda$ is the wavelength at the carrier frequency and $\sigma_{{\rm rcs}, p}$ and $d_{p}$ are the radar cross section (RCS) and range of the $p$-th scatter point.

Assuming a single \ac{LoS} path, with parameters $(\nu_0,\tau_0,\phi_0)$, between the \ac{Tx} and the \ac{UE} antenna, the downlink (communication-) channel is described by the impulse response 
\begin{align}\label{eq:ComChannel}
\Hm_C (t, \tau) =  \rho_0 \bv(\theta_0) \av^\herm (\phi_0)\delta(\tau-\tau_0) e^{j(2\pi \nu_0)t}\,,
\end{align}
where $\bv(\theta)$ denotes the array response of the \ac{Rx} at the user end.  $\rho_{0}$ is the communication channel gain given by $ |\rho_{0}|^{2} = \frac{\lambda^2 }{(4\pi)^2 (d_0)^2}$, where a path loss exponent of $2$ is considered, which is typical for mmWave LOS outdoor urban and rural scenarios.
%%%%%%%%%%%%%%%%%%%%%%%%%%%%%%%%%%%%%%%%%%%%%%%%%%%%%%%%%%%%%%%%%%%%%%%%%%%%%
\subsection{Signaling Scheme}
    We consider \ac{OFDM} as it is one of the standardized waveforms for \ac{mmWave} systems and \ac{ISAC} applications. By setting $\Delta f$ as the subcarrier spacing, $T_{\rm cp}$ the cyclic prefix duration, and $T_0 \eqdef 1/\Delta f + T_{\rm cp}$ as the total symbol duration including cyclic prefix, and considering $N$ symbols and $M$ subcarriers, the transmitted \ac{OFDM} frame is given by
    \begin{align}\label{eq:OFDM_tx}
        \sv(t) = \fv\sum_{n=0}^{N-1}\sum_{m=0}^{M-1}\zeta[n, m]p_{n, m}(t),
    \end{align}
    where $\zeta[n, m]$ is the $n$-th \textit{data} symbol intended to the user sent over subcarrier $m$, and $p(t)$ is a pulse shaping filter. 
    At the radar receiver, a reduction matrix $\Um$, designed as described at the beginning of this section, is applied before sampling. After standard \ac{OFDM} processing (see e.g. \cite{sturm2011waveform}), the noisy sampled signal is given by
    \begin{align}
        \rv[n, m] = \sum_{p=0}^{P-1}h_p\Um^H\av(\phi_{p})\av^H(\phi_{p})\fv\tilde{\zeta}[n, m] + \Um^H\wv[n, m] \label{eq:rx_sample},
    \end{align}
    where we defined $\tilde{\zeta}[n, m] \eqdef \zeta[n, m]e^{j2\pi(n T_0\nu_{p} - m\Delta f\tau_{p})}$, and $\wv[n, m] \in \CC^{\Na}$ is white Gaussian noise with $\EE\left[\wv[n, m]\wv^\herm[n, m]\right] = \sigma_{n}^{2}\Id_{\Na}$.
%%%%%%%%%%%%%%%%%%%%%%%%%%%%%%%%%%%%%%%%%%%%%%%%%%%%%%%%%%%%%%%%%%%%%%%%%%%%%%%%%%%%%%%%%%%%%%%%%%%%

%%%%%%%%%%%%%%%%%%%%%%%%%%%%%%%%%%%%%%%%%%%%%%%%%%%%%%%%%%%%%%%%%%%%%%%%%%%%%%%%%%%%%%%%%%%%%%%%%%%
\subsection{Maximum Likelihood Parameter Estimation}\label{sec:Joint-Detection-Param-Estimation}
Denote the true value of parameters as $\mathring{\thetav} = \{\mathring{h}_p, \mathring{\nu}_p, \mathring{\tau}_p, \mathring{\phi}_p\}_{p=0}^{P-1}$ 
%and using $\thetav = \{h_p, \nu_p, \tau_p, \phi_p\}$ to express the arguments of the likelihood function, (I think this is not used...)
%As said before, assuming that $\mathring{\thetav}$ is constant over $B$ OTFS blocks, target detection and parameter estimation are performed from the $B$-block observation.  
. The received signal expression (\ref{eq:rx_sample}) is written in a compact form by blocking the $NM$ Doppler-delay signal components into $NM$-dimensional vectors, where the underline symbol indicates blocked quantities. We define $\Tm(\tau, \nu) \in \CC^{NM\times NM}$ as
\begin{multline}
    \Tm(\tau, \nu) = 
    {\rm diag}([1, \dots, e^{j2\pi nT_{\rm 0}\nu}, \dots, e^{j2\pi (N-1)T_{\rm 0}\nu}]^\trasp \\
    \otimes [1, \dots, e^{-j2\pi m\Delta f\tau}, \dots, e^{-j2\pi (M-1)\Delta f \tau}]^\trasp),
\end{multline}
where $\otimes$ indicates the Kronecker product. For each $b \in[B]$, the  effective channel matrix of dimension $\Nrf NM\times NM$ associated to scattering point $p$ is given as ${\Gm}_{b}(\mathring{\nu}_{p}, \mathring{\tau}_{p}, \mathring{\phi}_{p}) \eqdef \Tm(\mathring{\tau}_{p}, \mathring{\nu}_{p}) \otimes \Um_{b}^{H}\av(\mathring{\phi}_{p})\av^H(\mathring{\phi}_{p})\fv $. The received signal then takes the form 
\begin{align}
    \underline{\rv_{b}}&= \sum_{p=0}^{P-1}\bigg( \Tm(\mathring{\tau}_p, \mathring{\nu}_p) \otimes \mathring{h}_p \Um_b^\herm \av(\mathring{\phi}_p)\av^\herm(\mathring{\phi}_p) \fv\bigg) \underline{\zetav}_b + \wv_b \nonumber \\
    &~=~ \sum_{p=0}^{P-1} \mathring{h}_{p} \Gm_b(\mathring{\nu}_{p}, \mathring{\tau}_{p}, \mathring{\phi}_{p}) \underline{\zetav}_b + \wv_b .
   \label{eq:rx_blocked}
\end{align}

Given knowledge of the number of scattering points $P$, the \ac{ML} estimate of the set $\mathring{\thetav}$ can be obtained by solving
\begin{align}\label{eq:full_ML}
    \thetav_{\rm ML} = \underset{\{h_{p}, \nu_{p}, \tau_{p}, \phi_{p}\}_{p=0}^{P-1}\in\Gamma}{\arg\min}\sum_{b=0}^{B-1}\left\| \underline{\rv}_{b} - \sum_{p=0}^{P-1} h_{p} \Gm_b(\nu_{p}, \tau_{p}, \phi_{p}) \underline{\zetav}_b \right\|_2^2,
\end{align}
where the search space is $\Gamma\eqdef\CC^P\times\RR^{3P}$. However, solving \eqref{eq:full_ML} requires prohibitively large computations and knowledge of the number of scattering points, a requirement hardly met in practice. Therefore, we resort to an approximate method that evaluates a hypothesis test on a set of $(\nu, \tau, \phi)$ tuples belonging to a coarse grid $\Theta$. In particular, for a given $(\nu, \tau, \phi)$ tuple, we perform a log-likelihood ratio test between the likelihood of $[\underline{\rv}_{1}, \dots, \underline{\rv}_{B}]$ corresponding to a single scattering point at location $(h', \nu, \tau, \phi)$ and the noise-only likelihood. The value $h'$ is chosen as the maximum likelihood estimate of the channel coefficient given that a scattering point is present at $(\nu, \tau, \phi)$, which can be evaluated in close form and is given by
\begin{align}
    h'(\nu, \tau, \phi) = \frac{\sum_{b=0}^{B-1}\underline{\zetav}_{b}^\herm \underline{\Gm}_{b}(\nu, \tau, \phi)^\herm\underline{\rv}_{b}}{\sum_{b=0}^{B-1}\|\underline{\Gm}_{b}(\nu, \tau, \phi)\underline{\zetav}_{b}\|_2^2}.
\end{align}
Using this value, the (generalized) log-likelihood ratio between the two hypotheses is given by
\begin{align}
    \ell(\nu,\tau,\phi) = \frac{\left | \sum_{b=1}^B \underline{\rv}_b^\herm \underline{\Gm}_b \underline{\zetav}_b \right |^2}{\sum_{b=1}^B \| \underline{\Gm}_b \underline{\zetav}_b \|_2^2}. \label{log-likelihood2S}
\end{align}
For every point in $\Theta$, the statistic \eqref{log-likelihood2S} is compared with a threshold, resulting in the generalized likelihood ratio test
\begin{equation} \label{GLRT}
    \ell(\nu,\tau,\phi)  \underset{\Hc_0}{\overset{\Hc_1}{\gtrless}} T_r \qquad (\nu, \tau, \phi) \in \Theta,
\end{equation}
where the threshold $T$ is chosen at each grid point by using the \ac{OS-CFAR} technique described in \cite{MultiDimCFAR}. More advanced techniques such as \ac{SIC} could instead be applied, at the cost of a more computationally expensive processing chain. However, as will be shown in Section \ref{sec:numeric_res}, a simple grid-based estimation and thresholding approach as proposed in this section yields excellent results for the considered \ac{ISAC} \ac{BT} problem.

%%%%%%%%%%%%%%%%%%%%%%%%%%%%%%%%%%%%%%%%%%%%%%%%%%%%%%%%%%%%%%%%%%%%%%%%%%%%%
% \begin{figure}[h]
% \centering
% \hspace{-1.4cm}

% \includegraphics[scale=0.6]{EuMW_Paper_LaTeX_Template_A4_V4/figures/LRR.pdf}
% \caption{Schematic}
% \label{fig:frame_strc}
% \end{figure}

%%%%%%%%%%%%%%%%%%%%%%%%%%%%%%%%%%%%%%%%%%%%%%%%%%%%%%%%%%%%%%%%%%%%%%%%%%%%
\section{Tracking for extended targets}\label{BF_arch}
In this section, we propose a scheme that uses the estimation methods described in Section \ref{sec:Joint-Detection-Param-Estimation} to track the user over time, in order to shift the transmit beam and appropriately select the set of beams at the receiver. As we discuss tracking over time, we introduce an index $t$ to the variables of interest, which indicates the time slot. 

As a first step, since the number of bins passing the \ac{OS-CFAR} test may vary from slot to slot, we summarize them into a single point in space, which subsequently becomes the magnitude we aim to track. In order to obtain such a point, we consider the weighted average of the locations that passed the \ac{OS-CFAR} test, weighted by their (appropriately normalized) log-likelihood ratios computed as shown in \eqref{log-likelihood2S}. This operation can be understood as obtaining the \textit{center-of-mass} of our estimate and effectively rejects outliers that pass the \ac{OS-CFAR} test but whose log-likelihood ratio is small. Notice that this applies only to the single target model considered in this paper. In the case of multiple extended targets, a clustering stage such as DBSCAN \cite{Kellner} would be performed after the \ac{OS-CFAR} test, and the weighted center would be obtained for each target separately.

Since we do not assume knowledge of any specific mobility model, we consider the method proposed in \cite{LiuLetter3}, which provides a simple predictor provided that the user acceleration varies in a much larger time scale than the interval between measurements. In particular, the kinematic equations for the $x$-coordinate (holding identically for the $y$-coordinate) considering three sampling epochs  are
 \begin{align}\label{eq:kinematics}
    \begin{cases}
        &x_{t+1} - x_{t} = v_{x, t}\Delta T + a_{x, t}\Delta T^{2}/2,\\
        &x_{t} - x_{t-1} = v_{x, t-1}\Delta T + a_{x, t-1}\Delta T^{2}/2,\\
        &x_{t-1} - x_{t-2} = v_{x, t-2}\Delta T + a_{x, t-2}\Delta T^{2}/2,\\
        &v_{x, t} = v_{x, t-1} + a_{x, t-1}\Delta T,\\
        &v_{x, t-1} = v_{x, t-2} + a_{x, t-2}\Delta T
    \end{cases},
\end{align}
where $(x_t, v_{x, t}, a_{x, t})$ represent the position, speed and acceleration in the $x$-axis at time $t$, and $\Delta T$ indicates the interval between measurements. If the acceleration in the last three measurement epochs is assumed constant, and measured values are used in lieu of ground truth ones, we arrive at the predictor
\begin{align}
    \hat{x}_{t+1} \approx 3\check{x}_{t} - 3\check{x}_{t-1} + \check{x}_{t-2}, \qquad \hat{y}_{t+1} \approx 3\check{y}_{t} - 3\check{y}_{t-1} + \check{y}_{t-2},
\end{align}
where $\check{x}_{t}$ indicates an estimated value and $\hat{x}_{t}$ a predicted one. From this values, it is easy to see that the predicted angle can be recovered as
\begin{align}\label{eq:pred_angle}
    \hat{\phi}_{t+1} = \tan^{-1}\left(\frac{3\check{x}_{t} - 3\check{x}_{t-1} + \check{x}_{t-2}}{3\check{y}_{t}-3\check{y}_{t-1}+\check{y}_{t-2}}\right).
\end{align}

The predicted angle \eqref{eq:pred_angle} will dictate the direction towards which the \ac{Tx} beamforming vector $\fv_{t+1}$ will point at time $t+1$. However, to ensure that the full target is covered by the \ac{Tx} beam, we follow a similar approach as \cite{Caire_extended} and consider beams with different widths. In particular, we focus on a simple method that uses the predicted distance to the target to choose one from a finite set of beamwidths. More advanced techniques for dynamic beamwidth adaptation are an interesting topic of research but are out of the scope of this paper.

% \section{Adaptive Transmit Beamforming}\label{sec:variable_width}
% Taking a similar approach as \cite{Caire_extended}, in this section, we investigate the effects of an adaptive \ac{Tx} beamwidth on the system communication performance.

%  \begin{figure}
% 	\centering
% 	%\hspace{3cm}
%     \hspace*{-1em} 
% 	\includegraphics[width=0.99\linewidth]{EuMW_Paper_LaTeX_Template_A4_V4/figures/multi_block_VBW.pdf}
% 	\caption{Transmit beamformers with adaptive width.}
% 	\label{fig:system_scheme}
% \end{figure} 

\section{Numerical Results}\label{sec:numeric_res}

{% <-- We enclose the table in a group so that any redefinitions
%% are automatically undone at the end of the group.
%
% \setlength{\tabcolsep}{2mm}%
% \renewcommand{\arraystretch}{1.2}% for the vertical padding of table cells
% \newcommand{\CPcolumnonewidth}{not used}%
% \newcommand{\CPcolumntwowidth}{21mm}%
% \newcommand{\CPcolumnthreewidth}{12mm}%
% \newcommand{\CPcolumnfourwidth}{33mm}%
% \begin{table}[H]
% \caption{System parameters considered in the simulations.}
% \small% EuMW: need this to get the 9pt text size in table cells % TODO is correct ?
% \centering
% 	\begin{tabular}{|c|c|}
% 		\hline
% 		$N=100, M=4$ &  Tx-FoV = $15^{\circ}$\\ \hline
% 		$f_c=90.0$ [GHz],~$W=160$ [MHz] & Interval = x sec \\ \hline
% % 			$\nu_{\mathrm{grid}}\simeq 24.4$ [KHz] $\equiv$ 40.6 [m/s] & $\tau_{\mathrm{grid}}= 1e-8$ [s] $\equiv$ 6 [m] \\ \hline
% % 		$v_{\mathrm{res}}\simeq440$ [km/h] & $r_{\mathrm{res}}\simeq1$ [m]\\ \hline
% % 		$v_{\mathrm{max}}=N\cdot v_{\mathrm{res}}$ & $r_{\mathrm{max}}=M\cdot r_{\mathrm{res}}$ \\ \hline
% 		$P_{\rm tx}=26$ [dBm] & $\sigma_{\mathrm{rcs}}=20$ [dBsm] \\ \hline
% 		Noise PSD $N_{\rm 0} = 2\cdot10^{-21}$ [W/Hz]  &  P$_{\rm fa}$ $=10^{-4}$\\ \hline
% 		$B=4$ & $\Na=64$,~$\Nrf=4$ \\ \hline
% 	\end{tabular}
%     \label{tab:sys_parameters}
% %\vspace{-\baselineskip}% remove one line of space below this table
% \end{table}
% }% end of group enclosing the table

\setlength{\tabcolsep}{2mm}%
\renewcommand{\arraystretch}{1.2}% for the vertical padding of table cells
\newcommand{\CPcolumnonewidth}{not used}%
\newcommand{\CPcolumntwowidth}{21mm}%
\newcommand{\CPcolumnthreewidth}{12mm}%
\newcommand{\CPcolumnfourwidth}{33mm}%
\begin{table}[H]
\caption{System parameters considered in the simulations.}
\small% EuMW: need this to get the 9pt text size in table cells % TODO is correct ?
\centering
	\begin{tabular}{|c|c|}
		\hline
		$N=4, M=100$ &  Tx-FoV $\in\{7, 10, 15, 20\} [^{\circ}]$\\ \hline
		$f_c=90.0$ [GHz] & $W=160$ [MHz] \\ \hline
        $\Delta f=1.6$ [MHz] & $T_{\rm cp} = 1/6 \cdot 1/\Delta f \approx$ \SI{0.1}{\micro\second}\\ \hline
% 			$\nu_{\mathrm{grid}}\simeq 24.4$ [KHz] $\equiv$ 40.6 [m/s] & $\tau_{\mathrm{grid}}= 1e-8$ [s] $\equiv$ 6 [m] \\ \hline
% 		$v_{\mathrm{res}}\simeq440$ [km/h] & $r_{\mathrm{res}}\simeq1$ [m]\\ \hline
% 		$v_{\mathrm{max}}=N\cdot v_{\mathrm{res}}$ & $r_{\mathrm{max}}=M\cdot r_{\mathrm{res}}$ \\ \hline
		$P_{\rm tx}=16$ [dBm] & $\sigma_{\mathrm{rcs}}=20$ [dBsm] \\ \hline
		  $\Delta T \in \{100, 200\}$ [ms] &  P$_{\rm fa}$ $=10^{-4}$\\ \hline
		$B=4$ & $\Na=64$,~$\Nrf=4$ \\ \hline
        \multicolumn{2}{|c|}{Noise PSD $N_{\rm 0} = 2\cdot10^{-21}$ [W/Hz]} \\ \hline
	\end{tabular}
    \label{tab:sys_parameters}
%\vspace{-\baselineskip}% remove one line of space below this table
\end{table}
}% end of group enclosing the table

We test the proposed estimation and tracking approach on a set of trajectories that reflect a realistic urban scenario. At each simulation, the path taken as well as the speed and acceleration at each sampling time follow a random process. Therefore, in order to meaningfully present our results, we average performance at each location over all generated trajectories. The relevant system parameters considered for the simulations are summarized in Table \ref{tab:sys_parameters}. Notice that the interval between measurements $\Delta T$ is much larger than the duration of an \ac{OFDM} block $BN(1/\Delta f + T_{\rm cp})$ in order to enable multi-target tracking through time division and/or reduce the duty cycle of the receiver to save energy.

\begin{figure}[!ht]
    \centering
    \includegraphics[width=0.48\textwidth]{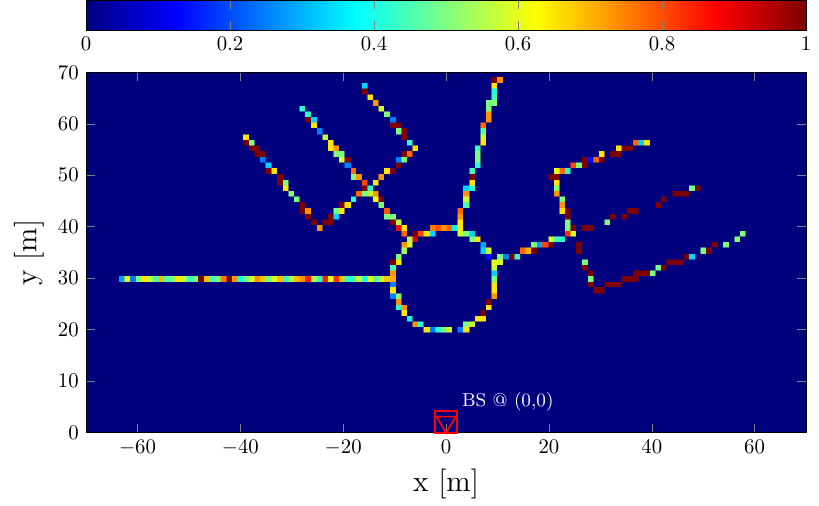}
    \caption{Probability that all scattering points are covered by the mainlobe of the transmit beam as a function of vehicle position along the trajectory.}
    \label{fig:prob_full_covered_vbw}
\end{figure}

As performance metrics, we first consider the probability of the communication signal being received at the user. Since generally the location of the antenna in the vehicle is unknown, we focus on the pessimistic case in which the user is only assumed to be able to decode the signal if all its scattering points fall within the main lobe of the transmitted beam. Fig.~\ref{fig:prob_full_covered_vbw} shows how our proposed estimator and tracker are accurate enough to fully cover the user at most locations with high probability. 
% It should be noted that since we use a constant width {\RED Not anymore! To update} transmit beam, at the closer distances the (lateral) vehicle positioning leads to a large angular spread as observed by the radar and therefore some sections of the vehicle might remain uncovered. Using different transmit beamwidths depending on the estimated position of the user is a subject of further research.

\begin{figure}[!ht]
    \centering
    \includegraphics[width=0.49\textwidth]{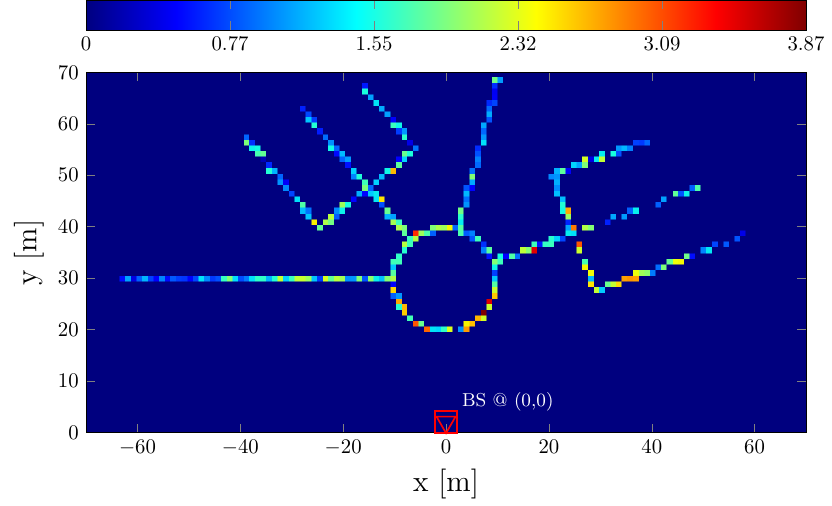}
    \caption{Achievable spectral efficiency in [bps/Hz] for the downlink channel as a function of the position.}
    \label{fig:spectral_efficiency_vbw}
\end{figure}

We continue by evaluating performance from a communications point of view. In particular, we study the achievable downlink spectral efficiency. For that, we consider the channel \eqref{eq:ComChannel} where, for simplicity, the \ac{UE} is assumed to receive through a single antenna\footnote{This does not incur any loss of generality since the methods presented here are not influenced by the specific antenna architecture of the \ac{UE}.}. Then, the achievable spectral efficiency is computed as
\begin{align}
    {\rm SE} = \log_{2}\left(1 + \frac{P_{\rm tx}|\av^\herm(\phi)\fv|^{2}}{N_{\rm 0}W}\left(\frac{\lambda}{4\pi d}\right)^2\right),
\end{align}
with the different parameters taking the values specified in Table \ref{tab:sys_parameters}. Similar to before, we only consider the term $|\av^\herm(\phi)\fv|$ to be greater than zero whenever all the scattering points fall within the main lobe of the beam generated by $\fv$. The resulting performance is shown in Fig.~\ref{fig:spectral_efficiency_vbw}. This figure shows how information can be transmitted to the user as it moves throughout a trajectory with an achievable spectral efficiency of at least 2 bps/Hz, and often significantly higher.

\begin{figure}
    \centering
    
     \begin{subfigure}{\linewidth}
    	\centering
    	%\hspace{3cm}
        \hspace*{-1em} 
    	\includegraphics[width=0.99\linewidth]{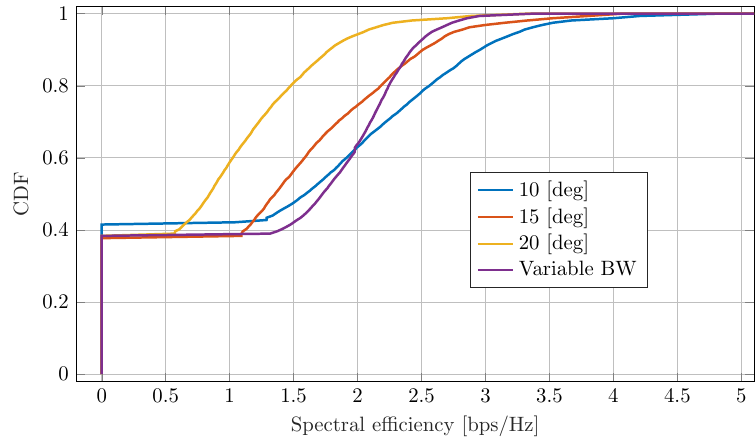}
    	\caption{$\Delta T = 100$ [ms]}
    	\label{fig:cdf_100}
    \end{subfigure} 
    \hfill
    
     \begin{subfigure}{\linewidth}
    	\centering
    	%\hspace{3cm}
        \hspace*{-1em} 
    	\includegraphics[width=0.99\linewidth]{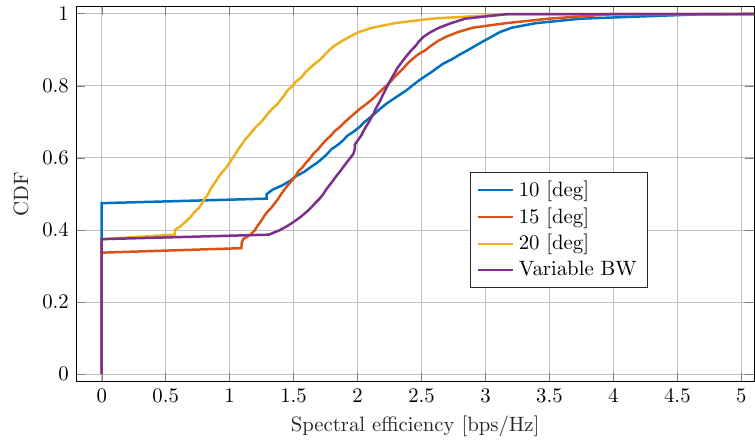}
    	\caption{$\Delta T = 200$ [ms]}
    	\label{fig:cdf_200}
    \end{subfigure} 
    \caption{\ac{CDF} of the achievable spectral efficiency for variable \ac{Tx} beamwidth and some fixed width beamformers, when the interval between measurements is \ref{fig:cdf_100} $\Delta T=100$ [ms] and \ref{fig:cdf_200} $\Delta T=200$ [ms].}
    \label{fig:SE_fixed_vs_variable}
\end{figure}

Finally, we study the effect of adapting the \ac{Tx} beamwidth based on the predicted distance to the target, and compare it with a simpler implementation in which the shape of the beam is kept constant. For the sake of clarity, we now aggregate the results at all positions and present performance in terms of the empirical \ac{CDF}, as shown in Fig.~\ref{fig:SE_fixed_vs_variable}  for two different values of the interval between measurements. As expected, the maximum achievable rate obtained by the variable beamwidth scheme lies in between that of the method that always uses the widest beam and the one that always uses the narrowest. The figure also shows how lower beamwidth transmitters are more sensitive to the interval between measurements than their wider or adaptive counterparts, as can be inferred from the difference in the step size at the leftmost side of the curve. The steepness of the transition in the variable beamwidth case demonstrates stable performance, concentrating most of the distribution mass around a given achievable spectral efficiency value. Such a feature is beneficial to simplify rate adaptation in practical systems. As a further note, it should be mentioned that the inter-measurement interval $\Delta T$, is a system design parameter that implies a tradeoff between energy efficiency and tracking performance and can be tuned to meet some minimal requirements.

%\begin{figure}[!ht]
%    \centering
%    \includegraphics[width=0.45\textwidth]{EuMW_Paper_LaTeX_Template_A4_V4/figures/spectral_efficiency.png}
%    \caption{Achievable spectral efficiency for the downlink channel as a function of the position.}
%    \label{fig:spectral_efficiency}
%\end{figure}

\section{Conclusion}
In this work, an efficient target detection and parameter estimation, and beam tracking framework was developed for \ac{OFDM} \ac{ISAC} systems. The proposed framework is based on an \ac{HDA} transceiver and a simple tracking equation. An important aspect of the proposed scheme is the fact that the communication component of the system is not compromised since the \ac{OFDM} symbols used for sensing contain no pilots or sensing-specific waveforms. Moreover, the validation of our results in complicated road geometries suggests that the proposed method can cope well in general realistic mobility scenarios. 

%%%%%%%%%%%%%%%%%%%%%%%%%%%%%%%%%%%%%%%%%%%%%%%%%%%%%%%%%%%%%%%%%%%%%%%%%%%%%

\section*{Acknowledgment}
The authors would like to acknowledge the financial support by the Federal Ministry of Education and Research of Germany in the program of “Souverän. Digital. Vernetzt.” Joint project 6G-RIC, project identification number: 16KISK030. 

%%%%%%%%%%%%%%%%%%%%%%%%%%%%%%%%%%%%%%%%%%%%%%%%%%%%%%%%%%%%%%%%%%%%%%%%%%%%%

\bibliographystyle{IEEEtran}

\bibliography{IEEEabrv,IEEEexample}

\end{document}

%% file: macros.tex
\DeclareMathAlphabet{\pazocal}{OMS}{zplm}{m}{n}

\newfont{\bbb}{msbm10 scaled 700}

\newfont{\bb}{msbm10 scaled 1100}
\newcommand{\CC}{\mbox{\bb C}}

\newcommand{\RR}{\mbox{\bb R}}

\newcommand{\EE}{\mbox{\bb E}}

\newcommand{\av}{{\bf a}}
\newcommand{\bv}{{\bf b}}

\newcommand{\fv}{{\bf f}}

\newcommand{\rv}{{\bf r}}
\newcommand{\sv}{{\bf s}}

\newcommand{\uv}{{\bf u}}
\newcommand{\wv}{{\bf w}}

\newcommand{\Fm}{{\bf F}}
\newcommand{\Gm}{{\bf G}}
\newcommand{\Hm}{{\bf H}}
\newcommand{\Id}{{\bf I}}

\newcommand{\Tm}{{\bf T}}
\newcommand{\Um}{{\bf U}}

\newcommand{\Hc}{{\cal H}}

\newcommand{\zetav}{\hbox{\boldmath$\zeta$}}

\newcommand{\thetav}{\hbox{\boldmath$\theta$}}

\renewcommand{\arg}{{\hbox{arg}}}

\newcommand{\eqdef}{\stackrel{\Delta}{=}}

\newcommand{\herm}{{\sf H}}
\newcommand{\trasp}{{\sf T}}

\newcommand{\Na}{N_{\rm a}}
\newcommand{\Nrf}{N_{\rm rf}}

\newcommand{\Ns}{N_{\rm s}}